\documentclass[doublecol,figures]{epl2}
\usepackage[latin1]{inputenc}
\usepackage{color}
\usepackage{soul}

\title{Glassy Transition in the Vortex Lattice of Ba(Fe$_{0.93}$Rh$_{0.07}$)$_2$As$_2$\\ superconductor probed by NMR and ac-susceptibility}

\shorttitle{Glassy Transition in the Vortex Lattice, in an iron-based superconductor}

\author{L. Bossoni\inst{1,2} \and P. Carretta\inst{1} \and M. Horvati\'c\inst{3} \and M. Corti\inst{1} \and A. Thaler\inst{4} \and P. C. Canfield\inst{4}}

\institute{
\inst{1} Department of Physics, University of
Pavia-CNISM -  I-27100 Pavia, Italy\\
\inst{2} Department of Physics ``E. Amaldi,'' University of Roma Tre - CNISM - I - 00146 Roma, Italy\\
\inst{3} Laboratoire National des Champs Magn\'etiques Intenses, LNCMI-CNRS (UPR3228),
UJF, UPS and INSA, BP 166, 38042 Grenoble Cedex 9, France\\
\inst{4} Ames Laboratory USDOE and Department of Physics and
Astronomy, Iowa State University -  Ames, IA 50011, USA
}


\abstract{ By using Nuclear Magnetic Resonance and
ac-susceptibility, the characteristic correlation times for the
vortex dynamics, in an iron-based superconductor, have been
derived. Upon cooling, the vortex dynamics displays a crossover
consistent with a vortex glass transition. The correlation
times, in the fast motions regime, merge onto a universal curve
which is fit by the Vogel-Fulcher law, rather than by an
Arrhenius law. Moreover, the pinning barrier shows a weak
dependence on the magnetic field which can be heuristically
justified within a fragile glass scenario. In addition, the glass
freezing temperatures obtained by the two techniques merge onto
the de Almeida-Thouless line. Finally the phase diagram for
the mixed phase has been derived.}

\begin{document}

\maketitle

\section{Introduction}

The discovery of iron-based superconductors has brought to a
renewed interest in the study of the mixed phase of type-II
superconductors and of the nature of the frozen vortex state.
In this regard, the effect of quenched disorder on the flux lines
lattice (FLL) properties has to be considered, together with the
onset of a vortex glassy state\cite{Huse1992,Blatter1996,Marchetti1999,Fisher1989}. In the
case of weakly anisotropic superconductors, such as the
122 family of iron-pnictides, the high temperature ($T$) region of
the phase diagram is dominated, just below the transition
temperature $T_c$, by a high mobility state of vortex lines\cite{book0}. On
the other hand, the nature of the low temperature vortex lattice
(VL) dynamics is still debated, since it involves the complex
interplay among the pinning forces, the intervortex repulsion/attraction \cite{Blatter1994} and
the thermal excitations. In a recent theoretical work\cite{Li2003}
the free energy of the vortex matter was explicitly calculated for a three dimensional (3D)
superconductor, in presence of quenched
disorder. Two transitions were found: a first-order melting,
at which the quasi-long-range order is destroyed, and a
second-order glassy transition. Such double transition has been recently
experimentally observed in cuprates\cite{Beidenkopf}. By taking
advantage of the work carried out in cuprates, it is worth
extending the investigation of the VL to the iron-based
superconductors, keeping in mind the different structural and physical properties.

ac-susceptibility and Nuclear Magnetic Resonance (NMR)
spin-lattice relaxation, are excellent probes to investigate the frequency
($\nu$) and magnetic field ($H$) dependence of the correlation
times ($\tau_c$) of vortex motions, and their joint employ is
particularly effective. In both cases the VL excitations are
probed in the radio-frequency range but, while the former
technique is sensitive to dissipative/dispersive mechanisms
taking place at wave-vector $\vec{q}\rightarrow0$, the latter is
sensitive to the $\vec{q}$-integrated dynamics.

In this paper, measurements of the real $\chi'$ and imaginary part
$\chi''$ of the ac-susceptibility are presented and discussed,
together with $^{75}$As NMR spin-lattice relaxation rate
$1/T_1$, in a single crystal of
Ba(Fe$_{0.93}$Rh$_{0.07}$)$_2$As$_2$ superconductor. The
temperature dependence of the vortex correlation times, extracted
both from the peaks in $1/T_1$ and $\chi''$, lead to an
interpretation of the motion, characteristic of a liquid-glassy
phase transition. It is noticed that the peaks in $1/T_1$
associated with the VL motions have been found only in a few
iron-pnictides.\cite{Laplace2009,Bossoni,Baek2009} This effect
cannot be ascribed to a coherence peak. In fact it has been
theoretically predicted \cite{Parker2008} that for an s$\pm$
wave superconductor, which is the most likely scenario in these
compounds, no divergence in the density of states (DOS) is
expected. Moreover, as shown in \cite{Bossoni}, even in the
case of s-wave pairing the exponential dependence of the
spin-lattice relaxation rate, at low temperature, is not
consistent with a superconducting gap of the order of meV.
\cite{Lester2010}

\section{Experimental details}
The ac-susceptibility measurements were carried out by means of a
Quantum Design MPMS-XL5 Squid ac-susceptometer. The sample was
mounted on the experimental setup, with both the static $H_{dc}$
and the oscillating $H_{ac}$ fields laying in the crystallographic
$ab$ plane. The sample dimensions, compared to the ones of the
experimental setup, allowed to measure the spin susceptibility for
the $\mathbf{H
\parallel ab}$ geometry only.  The ac-susceptibility measures the
energy dissipation due to vortex dynamics at wavevector $\vec{q}
\rightarrow 0$ and is complementary to NMR $1/T_1$ which, using
local probes, measures the $\vec q$-integrated dynamics.

A magnetic characterization of the sample was
previously carried out, to extract both $T_c(0) \sim 23$ K and the
irreversibility curve $T_{irr}(H)$. All the measurements were
performed in Field Cooled (FC) conditions, on a flat 0.8 x
5 x 7 mm$^3$ single crystal of
Ba(Fe$_{0.93}$Rh$_{0.07}$)$_2$As$_2$. The details regarding the
sample growth can be found in Ref. \cite{Bossoni}. During
the experiment the intensity of the oscillating magnetic field
was kept constant at $H_{ac}$ = 1.5 Oe, while the $dc$ field
intensity ranged from $H_{dc}$ = 500 Oe to 4.8 T, so that the
mixed phase regime could be explored. The \textit{ac} field frequency
range was $\nu$ = 37.5-1488 Hz.

\begin{figure}[h!]
    \onefigure[width=8.5cm]{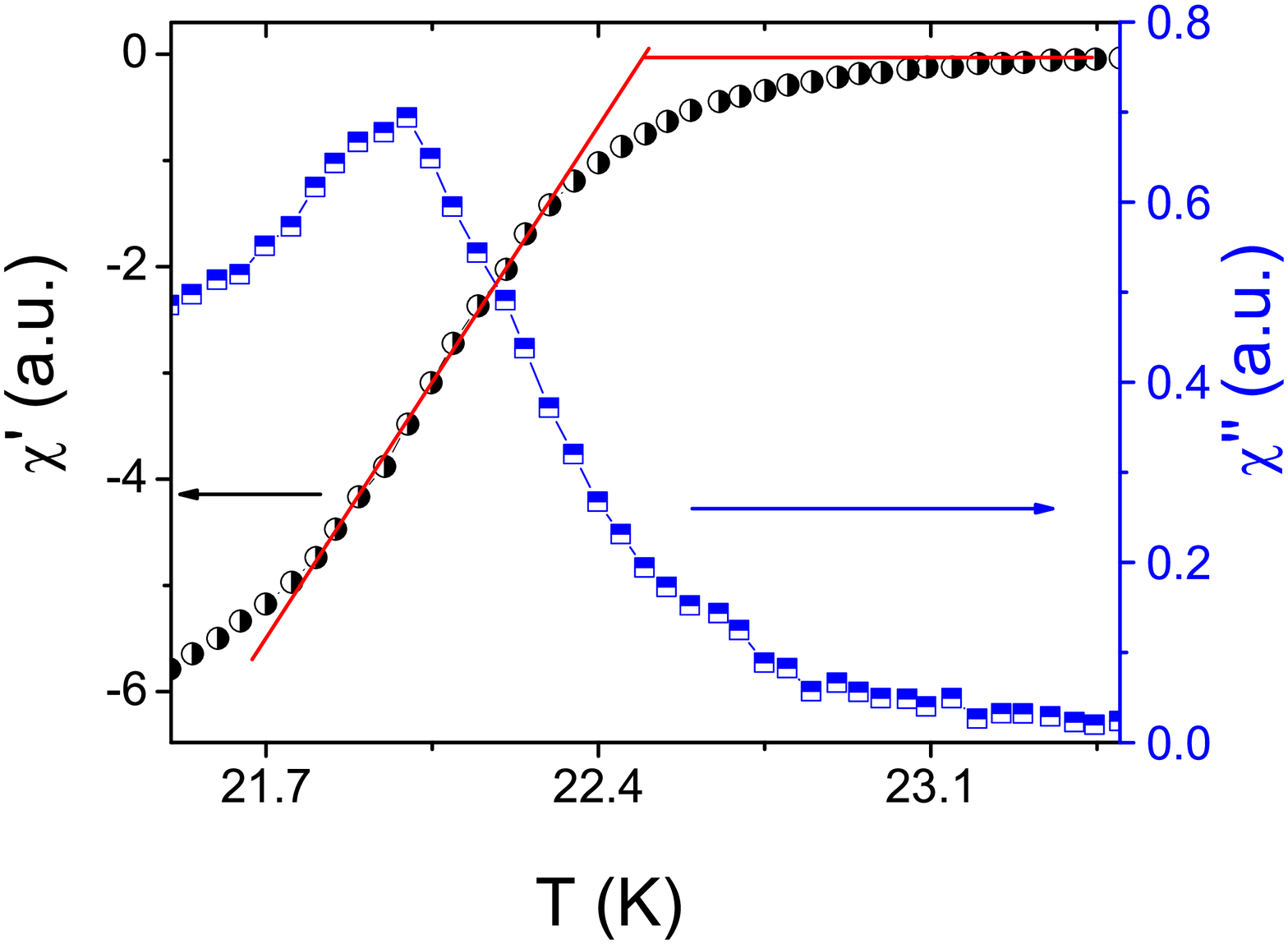}
    \caption{Temperature dependence of $\chi'$ (black circles), and $\chi''$ (blue squares) measured at $H=$ 500 Oe, and at $\nu=37.5$ Hz. Below the transition temperature marked by the drop in $\chi'$, the peak in $\chi''$ appears, pointing out that some dissipation  mechanism plays role in the mixed phase. The red lines are fits to determine $T_c$}
    \label{chi}
    \end{figure}

\begin{figure}[h!]
    \onefigure[width=8.5cm]{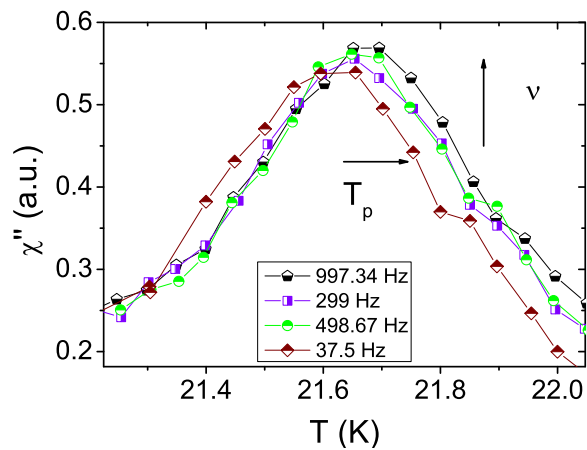}
    \caption{Frequency $\nu$ and temperature $T$ dependence of $\chi''$ peaks,
    measured at $H=$ 1 T. The peak temperature $T_p$ shifts towards
    higher temperature when increasing $\nu$.}
    \label{raw_chi}
    \end{figure}

\begin{figure}[h!]
    \onefigure[height=6cm]{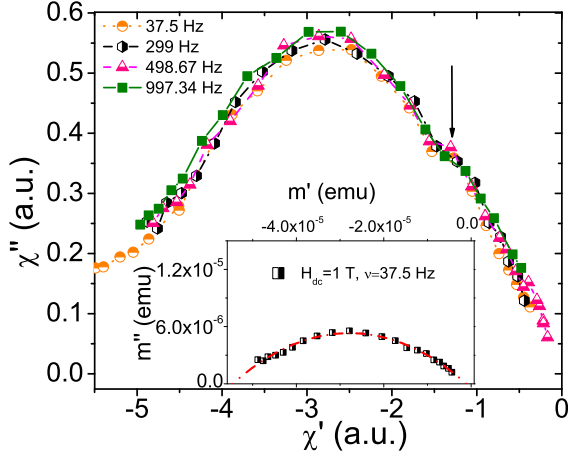}
    \caption{Cole-Cole plot, at
    $H=$ 1 T, with the temperature as implicit parameter. The absolute
    value of $\chi''$ increases slightly but systematically with
    increasing $\nu$, as expected for a vortex glass
    phase\cite{Adesso2004}. The arrow shows a kink probably due to a
    residual inter-grain contribution\cite{Polichetti}. The inset
    shows the Cole$-$Cole plot, and the dashed red line shows the fit (see text).}
     \label{ccp}
\end{figure}

Fig. \ref{chi} shows the real $\chi'$ and imaginary $\chi''$ part of the spin susceptibility.
 From the real part the transition temperature can be evaluated, whereas the peak in $\chi''$ provide information about the dissipative mechanisms occurring in the mixed phase.  
Fig. \ref{raw_chi} shows a peak in the imaginary part of
the ac-susceptibility $\chi''(T)$ and its evolution with the
ac-frequency: the reader can observe that the peak moves to higher
temperature when increasing the frequency. Fig. \ref{ccp} displays the
imaginary part of the spin susceptibility as a function of the
real part of the spin susceptibility, with the temperature as
implicit parameter. This curve is usually known as Cole$-$Cole
plot. The inset of the figure shows that the Cole$-$Cole plot lays
on a circle, whose center has a negative ordinate, pointing out
that there is a likely distribution of correlation times
$\tau_c$.\cite{Mognaschi}

\begin{figure}
\onefigure[width=8.5cm, keepaspectratio]{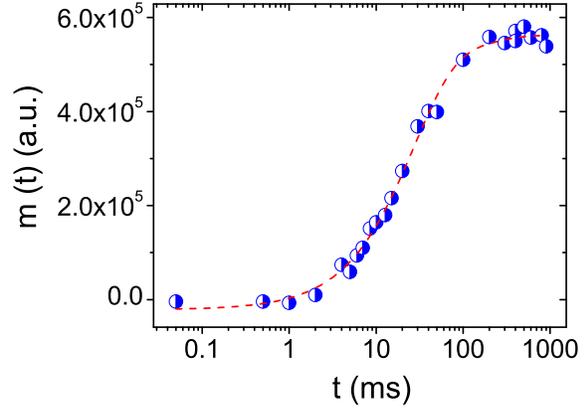}
\caption{The raw saturation recovery data at 8.5 T ($\mathbf{H}\parallel c$) and 16.5 K. The dashed red line is the best fit according to equation (1). From the best fit, the stretched exponent $r$ is found to be $r=0.93$.}
\label{recovery}
\end{figure}

The $^{75}$As ($I=3/2$) NMR measurements were performed on the
same single crystal, by using standard radio-frequency pulse sequences, on
irradiating and detecting only the central $m_I=-1/2
\longrightarrow 1/2$ line. The spin-lattice relaxation time $T_1$
was measured by means of a saturation recovery pulse sequence and the recovery of nuclear magnetization $m(t)$ was fit by the law:
\begin{equation}
1-m(t)=0.1 \exp(-t/T_1)^r+0.9 \exp(-6t/T_1)^r
\end{equation}

with $r\longrightarrow 1$, as expected for the central transition of a nuclear spin $I=3/2$. Fig. \ref{recovery} shows the raw saturation recovery data with the fitting function, at 8.5 T, in the mixed phase.

\begin{figure}
\onefigure[width=8.5cm, keepaspectratio]{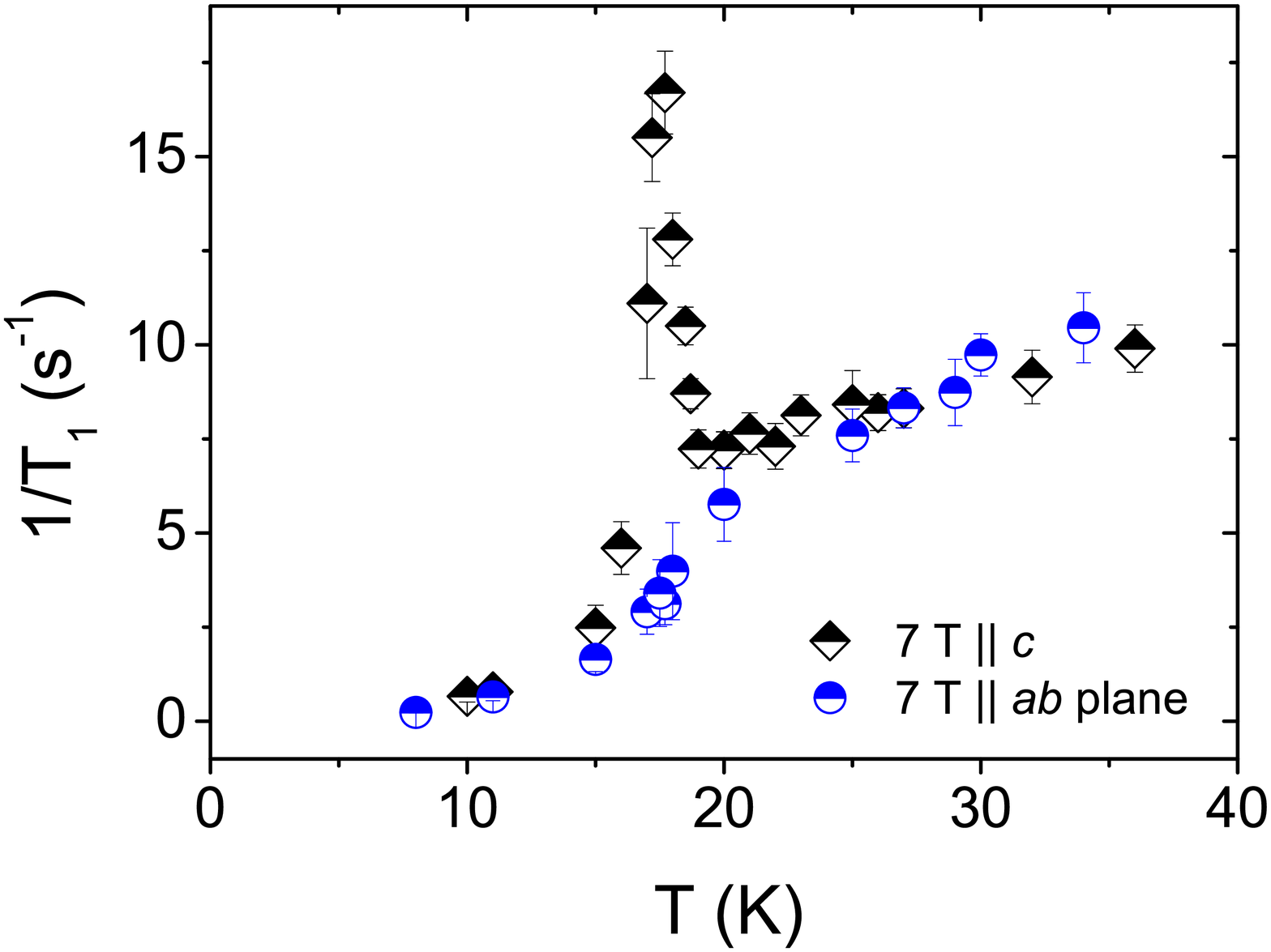}
\caption{The spin-lattice relaxation rate measured at 7 T, for
$\mathbf{H}\parallel c$ (black diamonds) and for $\mathbf{H}\perp
c$ (blue circles), rescaled by the hyperfine coupling ratio (see
Ref.\cite{Bossoni}). The peak is
visible solely in the longitudinal field geometry, while in the transverse geometry, $\mathbf{H}\perp
c$, only the electronic contribution $1/T_{1el}$ to
the nuclear spin-lattice relaxation is present.} \label{T1_7T}
\end{figure}

Earlier experiments, performed in the mixed
phase of the same compound, evidenced a peak in the spin-lattice
relaxation rate, whose position and amplitude depend on $H$
intensity \cite{Bossoni}.
As shown in Ref. \cite{Bossoni}, and reported in Fig. \ref{T1_7T} for clarity, the peaks are visible solely in $\mathbf{H} \parallel c$ geometry, as expected for a layered superconductor, where the nuclear relaxation is driven by the vortex dynamics.

The contribution of the vortex dynamics to the spin-lattice relaxation
rate $1/T_{1VL}$, for $\mathbf{H}\parallel c$ is found by subtracting from the raw data the
electronic contribution, estimated by extrapolating the data out
of the peak region. This procedure is in practice equivalent to
subtracting the data taken for $\mathbf{H} \perp
c$, properly rescaled by the hyperfine factors\cite{Bossoni}, since for that
orientation the nuclear spin-lattice relaxation is only given by
the electronic spin fluctuations (Fig. \ref{T1_7T}). Namely, one
has
\begin{equation}
\frac{1}{T_{1}}=\frac{1}{T_{1VL}}+\frac{1}{T_{1el}},
\end{equation}
where the first term is the vortex-lattice contribution, whilst the
latter is due to electronic spin fluctuations.

\begin{figure}
\includegraphics[width=8.5cm, keepaspectratio]{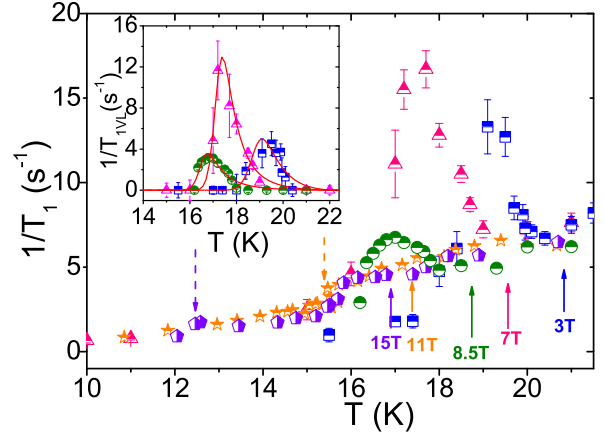}
\caption{The spin-lattice relaxation rate at different magnetic
field values: 3 T (blue squares), 7 T (pink triangles),  8.5 T
(green circles),  11 T (yellow stars),  and 15 T (purple
pentagons), is reported. All data refer to $\mathbf{H}\parallel c$
orientation. The arrows show the temperature of the detuning of
the NMR probe resonating circuit. The dashed arrows mark the position of the
kink in the $1/T_1$, when the field reaches 11 T and 15 T}. The inset shows the $1/T_{1VL}$
peak due to the VL dynamics, namely after the
subtraction of the electronic contribution. The solid lines are guides to-the-eye. 
\label{T1}
\end{figure}

Fig. \ref{T1} shows the relaxation rate $1/T_1$ versus
temperature, at all the applied fields. By increasing the field
intensity, the peak in $1/T_1$ shifts towards lower temperatures
and it is rapidly reduced above 7 T, so that only a small kink remains at
11 T and 15 T. The results of Fig. \ref{T1} show remarkable
similarities with earlier studies on oriented powders of the
cuprate YBa$_2$Cu$_4$O$_8$ (YBCO124)\cite{Corti1996,Rigamonti1998}.

Before concluding this section it is worth noting that, below the
temperature of the peak in $1/T_1$, a marked growth in the
acoustic ringing was observed at all field values, mostly for
$\mathbf{H} \parallel c$. In analogy with an ultra-sound
experiment\cite{Pankert}, this effect can be qualitatively
understood if one assumes that a magneto-acoustic coupling between
the VL and the crystal lattice is affecting the magnetic
susceptibility of the sample inside the NMR resonating coil.
If a liquid-glassy phase transition occurred in the the vortex
matter, the thermal excitation would not allow the vortices to
overcome the pinning barrier in the glassy state, so they would
transfer their elastic energy to the lattice, causing the latter
to vibrate.

\section{Discussion}

To gain insights into the VL dynamics, the maximum in $\chi''$ has
to be carefully analyzed. Just below the onset of
superconductivity, evidenced in the $\chi'$ curve, a sharp peak in
the $\chi''$ appears (Fig. \ref{chi}). The peak shifts
towards higher temperature, when increasing the frequency, thus
excluding the occurrence of the Bean critical
state\cite{Bean1962}. The peak in the imaginary part of the
susceptibility could be interpreted as the result of magnetic flux
reaching the center of the specimen and giving a resultant
magnetization $M$. This effect would be strongly dependent on the
sample geometry and on the intensity of the applied field. In such
a case, the peak temperature would not depend on the ac-frequency,
at variance with the experimental findings. However, to definitely
rule out the occurrence of Bean critical state, a further
analysis should be carried out, by varying the intensity of the
ac-field.

Another interpretation relies on the following consideration. The
peak in $\chi''$ is dominated by the onset of irreversible
behavior of the magnetization, occurring when the vortex lines are
thermally excited across the pinning barriers.\cite{Malozemoff}
This leads to a resonant absorption of energy for $2 \pi \nu
\tau_c \sim 1$. The absolute value of the imaginary part of the
susceptibility tends to increase, while increasing the frequency
$\nu$. The peak height of the Cole$-$Cole plot grows when increasing the
ac-frequency, and so does the peak in the imaginary part of the
spin susceptibility $\chi''(T)$. These findings are in agreement
with the numerical simulations of Adesso et al. as well as
with their experimental results on YBCO \cite{Adesso2004} and
LaFeAsO$_{0.9}$F$_{0.08}$\cite{Polichetti}. As a further
proof we point out that the same behavior has been recently
observed in the Cobalt optimally doped Ba122 \cite{Giacomo2012}.

To give quantitative information about the vortex dynamics,
the correlation time $\tau_c(H,T)$ dependence on the field H and
the temperature T, is extracted through the Debye relation
\begin{equation}
\chi'' \propto  \frac{\omega\tau_c}{1+(\omega\tau_c)^2},
\end{equation}

Given the sharpness of the peaks in $\chi''(T)$, and the previous consideration on the Cole-Cole plot, a small distribution of
correlation times may be assumed. Accordingly, the value of
$\tau_c$ shall be considered as an average value.

\begin{figure}[h!]
    \includegraphics[width=8.5cm,keepaspectratio]{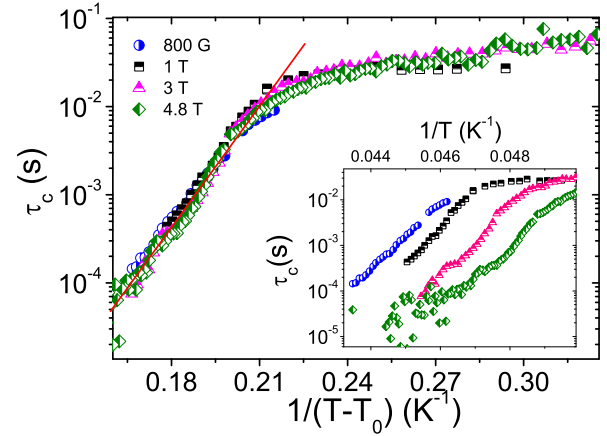}
    \caption{The correlation time for the $\vec{q}=0$ motions
    derived from ac-susceptibility data, reported as a function of
    $1/(T-T_0)$, is shown. The data nicely merge onto a universal curve
    that obeys the VF law (red line). In the inset the $\tau_c$ is
    reported as a function of $1/T$. The dynamical crossover is
    evident.}
    \label{tauc_susc}
\end{figure}

Fig. \ref{tauc_susc} shows a crossover occurring in $\tau_c$, during the slowing-down
of the FLL motion, upon cooling. Hence the Arrhenius law, typically employed in
cuprates\cite{Palstra} and in the
SmFeAsO$_{0.8}$F$_{0.2}$\cite{Giacomo2011}, fails in this system.
One can further realize that the Arrhenius law has to be abandoned,
by plotting $\nu$ versus the temperature of the peak, as recently
done in the optimally doped Ba(Fe$_{1-x}$Co$_x$)$_2$As$_2$
\cite{Giacomo2012}. Furthermore, the inset of Fig. \ref{tauc_susc}
shows that the curves have the
same slope. Then, if the data are properly shifted by a
temperature $T_0(H)$, they merge onto a universal curve that can
be fit by the Vogel-Fulcher (VF) law \cite{Rao}:
\begin{equation}
\tau_c(T,H)=\tau_0 \exp\left[\frac{U_{eff}}{T-T_0(H)} \right]
\end{equation}
The data can be fit by a nearly field-independent activation barrier ($U_{eff}= 120 \pm 20$ K), in contrast with certain
families of superconductors, where the barrier decreases as $1/H$ \cite{Palstra} (see next paragraph).


     \begin{figure}[h!]
  \includegraphics[width=8.5cm,keepaspectratio]{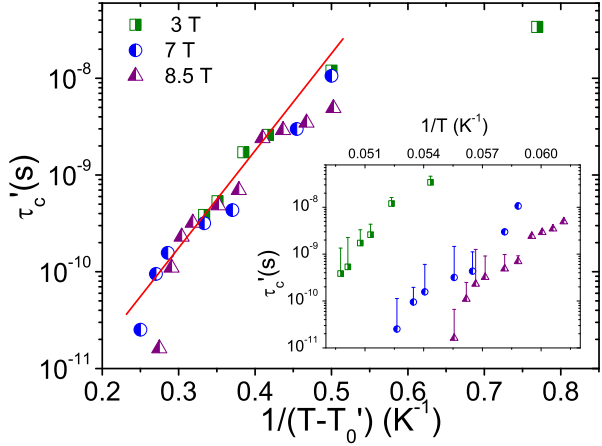}
 \caption{The $\vec{q}$-integrated correlation time $\tau_c'$ of the vortex
motions derived from $1/T_1$ is shown. The data are reported as a
function of $1/(T-T_0')$, and they merge onto a universal curve
that can be fit via the VF law (red line). The inset shows the
correlation time calculated by Eq. \ref{eq5}, before
correcting the temperature dependence. The NMR correlation times
show the same behavior as the ones extracted by ac-susceptibility,
hence showing an intrinsic common behavior.}  \label{tauc_NMR}
 \end{figure}


While the ac-susceptibility is sensitive to $q\rightarrow0$ excitations, the nuclear spin-lattice relaxation is driven by the $\vec{q}$-integrated dynamics of the thermally excited vortices. As shown in
Ref. \cite{Bossoni}, a correspondence between $1/T_1$ and the $\vec{q}$-integrated correlation time $\tau_c'$ can be found.
However here, instead of fitting the $1/T_1$ peaks, as already carried out on the same compound and on YBCO\cite{Corti1996},
the following equation is numerically solved, in order to extract the T-dependence of the correlation time $\tau_c'$:
\begin{equation}
\frac{1}{T_{1}}=\frac{(^{75}\gamma)^{2}}{2}\frac{\Phi_0^2 s^2 }{4
\pi \lambda_{ab}^4}<u^2>_{3D}
\frac{1}{\xi^2}\frac{1}{l_e^2\sqrt{3}}  \tau_c'\ln
\left[\frac{\tau_c'^{-2}+\omega_{L}^{2}}{\omega
_{L}^{2}}\right] \label{eq5},
\end{equation}

where $^{75}\gamma$ is the  $^{75}$As gyromagnetic ratio,
$<u^2>_{3D}$ the root mean-squared amplitude of the local field
fluctuations \cite{Glazmann1991}, $l_e$ is the inter-vortex spacing, $\omega_L$ is
the nuclear Larmor frequency, $s$ is the inter-planar spacing\cite{s},
$\lambda_{ab}$ is the London penetration depth when the magnetic field is oriented along the c axis, $\xi$ the coherence length and $\Phi_0$ the
magnetic flux quantum. This procedure is more reliable, because no
multiple fitting parameters are involved, and the crossover
between two dynamical regimes can be better evidenced.

One may wonder if the approach of Ref. \cite{Bossoni} is the
best one to describe $T_1$ in this weakly anisotropic system,
since it starts from the assumption that the 3D vortex lines can
be represented as stacks of 2D pancake vortices oscillating around
their equilibrium positions. To check whether the
dimensionality of the vortex is playing a role in the description
of the dynamics, a further analysis was done, by employing a
relation where the correlation time has a 3D isotropic
character\cite{HS1959}. No significant difference in the qualitative behavior of $\tau_c'$
was observed. Notice that the different absolute values of $\tau_c$ and $\tau_c'$ display a
dispersive behavior in the VL excitations.

As it was observed for $\tau_c$, also the $\tau_c'$ data merge onto a universal trend, after being
properly rescaled by $T_0'$ (Fig. \ref{tauc_NMR}). The VF fit value of the pinning barrier is $U_{eff}\sim 200 \pm 20$ K, not far from the
ac-susceptibility value. It is noticed that the energy barriers $U_{eff}$ are field
independent both for NMR and ac-susceptibility. This result,
together with the VF-like behavior, suggests the occurrence of a
glass state that, by resorting to the "glass
terminology"\cite{Varnik2002}, can be named \textit{fragile}
glass. The field independence of $U_{eff}$ can be justified by the
following qualitative consideration:
in case of weak pinning and high magnetic fields, namely in case of a high vortex density, the pinning energy distribution can be characterized by close meta-stable minima, in the bottom of spatially extended and deeper energy minima, as in the fragile glass scenario\cite{DeB,book}. Hence, upon varying the magnetic field strength, the VL can be rearranged within those meta-stable minima without having to overcome the high energy barrier. Accordingly, the barrier will correspond to an average energy distribution $<U>$,
determined solely by the quenched disorder, and not by the
magnetic field.

\begin{figure}[h!]
\includegraphics[height=6.2 cm,keepaspectratio]{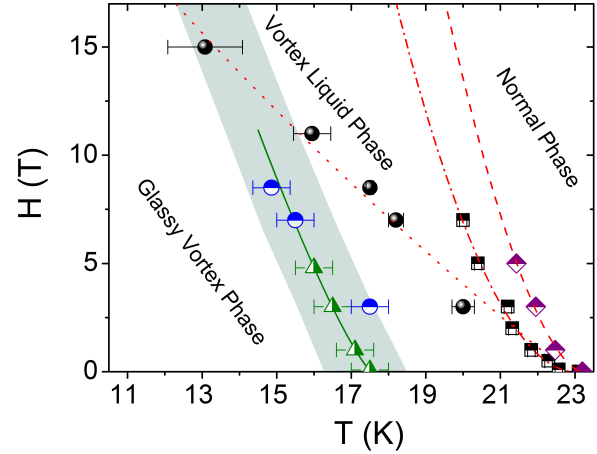}
\caption{The phase diagram for the mixed phase of the
Ba(Fe$_{0.93}$Rh$_{0.07}$)$_2$As$_2$ compound. The purple diamonds
 show the upper critical field $H_{c2}$ for
$\mathbf{H}\parallel c$, while the squares mark the
irreversibility temperature $T_{irr}$, for the static uniform
susceptibility, which coincides with the detuning temperature of
the NMR probe. The black circles mark the $T_{peak}$ line for the NMR
relaxation rate, namely  the temperature of the peak in $1/T_1$.
The green triangles represent $T_0$, while the blue half circles
represent $T_0'$. The green solid line is the de
Almeida-Thouless fit, while the other lines are guides to-the-eye.
The blue-grey zone marks the thermal disruption of the
vortices, and the crossover from slow dynamics (glassy state) to
fast dynamics (liquid state). } \label{PhD}
\end{figure}

When reporting $T_0$ and $T_0'$ in the
phase diagram, one finds a surprising result (see Fig.
\ref{PhD}): the temperatures estimated by the two techniques merge
onto the de Almeida-Thouless line\cite{deAlm}
$$
H=H_0[1-T_g(H)/T_g(0)]^{\gamma}.
$$
The exponent $\gamma \sim 1.5$ is in agreement with spin and superconductive
glasses\cite{Muller1987}.


 Before closing this section it is worth noting that both
$T_0$ and $T_0'$ lines intersect the $\mathbf{H}=0$ axes at a
temperature that is close to the zero field limit for the thermal
"disruption" of a vortex line. By assuming that a vortex line is
made by piling up many pancakes across the Fe-As planes, Clem
predicted that the thermal disruption of a vortex line would occur
at the Kosterlitz-Thouless transition temperature:\cite{Clem1991}
\begin{equation}
T_{cr} = \frac{\Phi_0^2 d}{32 \pi^2 k_B \lambda_{ab}(T)^2},
\end{equation}
where $d$ is the Fe-As layer thickness. The calculated value
of $T_{cr}$ for this compound is $\sim$ 18.5 K.\cite{nota}
This observation agrees with the picture of vortices passing from
a glass/solid phase to a liquid phase, namely a regime of poorly
correlated vortices.

\section{Conclusions}
NMR and ac-susceptibility are complementary techniques that
allow to shed light into the mixed phase of Ba122
superconductors. From both the techniques the correlation times
of the vortex dynamics were derived and it was shown that they obey
the VF law, in the fast motion regime. The behavior of $T_0$ and
$T_0'$ with the field, and the analysis of $\chi''$, unambiguously
show that a vortex glass phase is present in the low-T regime of
the phase diagram of the 122 family of iron-pnictides. $T_0$ and
$T_0'$ have been found to merge  onto the de Almeida Thouless
line. To gain further insights into the behavior of $U_{eff}$ with
the field, and on the predicted Kosterlitz-Thouless transition,
resistivity measurements will be performed.

\acknowledgments
We are indebted to A. Rigamonti and G. Jug for fruitful
discussions. We gratefully acknowledge C. Pernechele, M. Mazzani,
and R. DeRenzi for the support during the ac-susceptibility
measurements. We also acknowledge H. Mayaffre and M.-H. Julien of
LNCMI-Grenoble. This work was supported by Fondazione Cariplo
(research grant n.2011-0266) and by the European Commission
through the EuroMagNET II network (Contract No. 228043). Work done
in Ames Lab (PCC and AT) was supported by the U.S. Department of
Energy, Office of Basic Energy Science, Division of Materials
Sciences and Engineering. The research was performed at the Ames
Laboratory. Ames Laboratory is operated for the U.S. Department of
Energy by Iowa State University under Contract No.
DE-AC02-07CH11358.


\begin{thebibliography}{0}

\bibitem{Huse1992}
  \Name{D. A. Huse, M. P. A. Fisher \and D. S. Fisher}
 \REVIEW{Nature}{358}{553}{1992}.

 \bibitem{Blatter1996}
    \Name{G. Blatter, V. Geshkenbein, A. Larkin \and H. Nordborg}
    \REVIEW{Phys. Rev. B}{54}{72}{1996}.

\bibitem{Marchetti1999}
    \Name{M. C. Marchetti \and D. R. Nelson}
    \REVIEW{Phys. Rev. B}{59}{13642}{1999}.

\bibitem{Fisher1989}
    \Name{M. P. A. Fisher}
    \REVIEW{Phys. Rev. Lett.}{62}{1415}{1989}.

\bibitem{book0}
    \Name{E. Zeldov}
  \Book{100 Years of Superconductivity}
  \Editor{H. Rogalla \and P. H. Kes}
  \Publ{CRC Press, Boca Raton, FL}
  \Year{2012}
  \Page{228}.

\bibitem{Blatter1994}
    \Name{G Blatter, M. Y. Feigel'man, Y. B. Geshkenbein, A. I. Larkin, V. M. Vinokur}
    \REVIEW{Rev. Mod. Phys.}{66}{1125}{1994}.

\bibitem{Li2003}
    \Name{D. Li \and B. Rosenstein}
    \REVIEW{Phys. Rev. Lett.}{90}{167004}{2003}.

\bibitem{Beidenkopf}
    \Name{H. Beidenkopf, N. Avraham, Y. Myasoedov, H. Shtrikman, E. Zeldov, B. Rosenstein, E. H. Brandt \and T. Tamegai}
    \REVIEW{Phys. Rev. Lett.}{95}{257004}{2005}.

\bibitem{Laplace2009}
    \Name{Y. Laplace J. Bobroff, F. Rullier-Albenque, D. Colson, \and A. Forget}
    \REVIEW{ Phys. Rev. B}{80}{140501(R)}{2009}.

\bibitem{Bossoni}
    \Name{L. Bossoni, P. Carretta, A. Thaler \and P. C. Canfield}
    \REVIEW{ Phys. Rev. B }{85}{104525}{2012}.

\bibitem{Baek2009}
    \Name{S.-H. Baek, H. Lee, S. E. Brown, N. J. Curro, E. D. Bauer, F. Ronning, T. Park, and J. D. Thompson}
    \REVIEW{ Phys. Rev. Lett. }{102}{227601}{2009}.


\bibitem{Parker2008}
    \Name{D. Parker, O. V. Dolgov, M. M. Korshunov, A. A. Golubov, and I. I. Mazin}
    \REVIEW{Phys. Rev. B.}{78}{134524}{2008}.


\bibitem{Lester2010}
    \Name{C. Lester, Jiun-Haw Chu, J. G. Analytis, T. G. Perring, I. R. Fisher, and S. M. Hayden}
    \REVIEW{Phys. Rev. B.}{81}{064505}{2010}.

\bibitem{Mognaschi}
    \Name{E. R. Mosgnaschi and A. Rigamonti}
    \REVIEW{Phys. Rev. B}{14}{2005}{1976}.

\bibitem{Corti1996}
    \Name{M. Corti, J. Suh, F. Tabak, A. Rigamonti, F. Borsa, M. Xu \and B. Dabrowski}
    \REVIEW{Phys. Rev. B}{54}{9469}{1996}.

\bibitem{Rigamonti1998}
    \Name{A. Rigamonti, F. Borsa \and P.Carretta}
    \REVIEW{Rep. Prog. Phys.}{61}{1367}{1998}.


\bibitem{Pankert}
    \Name{J. Pankert, G. Marbach, A. Comberg, P. Lemmens, P. Fr\"{o}ning \and S. Ewert}
    \REVIEW{Phys. Rev. Lett.}{65}{3052}{1990}.

\bibitem{Bean1962}
    \Name{C. P. Bean}
    \REVIEW{Phys. Rev. Lett.}{8}{250}{1962}.

\bibitem{Malozemoff}
    \Name{A. P. Malozemoff, T. K. Worthingtin, Y. Yeshurun, F. Holtzberg \and P- H. Kes}
    \REVIEW{Phys. Rev. B(R)}{38}{7203}{1988}.

\bibitem{Adesso2004}
    \Name{M. G. Adesso, M. Polichetti \and S. Pace}
    \REVIEW{Physica C}{401}{196}{2004}.

\bibitem{Polichetti}
    \Name{M. Polichetti, M. G. Adesso, D. Zola, J. Luo, G. F. Chen, Z. Li, N. L. Wang, C. Noce \and S. Pace}
    \REVIEW{Phys. Rev. B}{78}{224523}{2008}.



\bibitem{Giacomo2012}
    \Name{G. Prando, M. Abdel-Hafiez, S. Aswartham, A.U.B. Wolter, S. Wurmehl \and B. Buchner}
    \REVIEW{arXiv:}{1207.2457v1}{2012}{unpublished}.
    \Name{G. Prando, P. Carretta, R. De Renzi, S. Sanna, H.-J. Grafe, S. Wurmehl, B. B\"{u}chner}
	\REVIEW{Phys. Rev. B} {85} {144522} {2012}.

\bibitem{Palstra}
    \Name{T. M. Palstra, B. Batlogg, R. B. van Dover, L. F. Schneemeyer \and J. V. Waszczak}
    \REVIEW{Phys. Rev. B}{41}{6621}{1990}.

\bibitem{Giacomo2011}
    \Name{G. Prando, P. Carretta, R. De Renzi, S. Sanna, A. Palenzona, M. Putti \and M. Tropeano}
    \REVIEW{Phys. Rev. B}{83}{174514}{2011}.
    

\bibitem{Rao}
    \Name{F. Rao, A. Cristanti \and F. Rotort}
    \REVIEW{Europhys. Lett.}{62}{869}{2003}.




\bibitem{Glazmann1991}
    \Name{L. I. Glazman \and A. E. Koshelev}
    \REVIEW{Phys. Rev. B}{43}{2835}{1991}.


\bibitem{s} {The used value for the inter-planar spacing is $s\sim 6 $ \AA.}

\bibitem{HS1959}
    \Name{L. C. Hebel \and C. P. Slichter}
    \REVIEW{Phys. Rev.}{113}{1504}{1959}.




\bibitem{Varnik2002}
    \Name{F. Varnik \and K. Binder}
    \REVIEW{Journ. of Chem. Physics}{117}{6336}{2002}.

\bibitem{DeB}
 \Name{P. G. Debenedetti \and F. H. Stillinger}
  \REVIEW{Nature}{410}{259}{2001}.

\bibitem{book}
 \Name{C. A. Angell}
  \Book{Insulating and semiconducting glasses, Series on Directions in Condensed Matter Physics}
  \Editor{P. Boolchand}
  \Vol{17}
  \Publ{World Scientific Publishing, Singapore}
  \Year{2000}
  \Page{18}.

\bibitem{deAlm}
    \Name{J. R. L. de Almeida \and D. J. Thouless}
    \REVIEW{J. Phys. A}{11}{983}{1978}.

\bibitem{Muller1987}
    \Name{K. A. M\"{u}ller, M. Takashige \and J. G. Bednorz}
    \REVIEW{Phys. Rev. Lett.}{58}{1143}{1987}.

\bibitem{Clem1991}
    \Name{J. R. Clem}
    \REVIEW{Phys. Rev. B}{43}{7837}{1991}.


\bibitem{nota} {For the estimate of $T_{KT}$, the used numerical values are the plane thickness $d=2.1$ \AA, and the bulk penetration depth $\lambda_{ab}= 240$ nm. Notice that $T_{KT}$ takes into account the T-dependence of $\lambda_{ab}$, according to the two-fluid model}.

\end{thebibliography}
\end{document}